**Dielectric transparency induced by hetero-phase oscillations near the phase transition of relaxor ferroelectrics**

J. Toulouse and R.K. Pattnaik*
Physics Department and Center for Advanced Materials, Lehigh University, Bethlehem, PA
L.A. Boatner
Center for Radiation Detection Materials and Systems, Oak Ridge National Laboratory, Oak Ridge, TN

ABSTRACT
We report the observation of a remarkable "transparency window" in the dielectric resonant absorption spectrum of the relaxor ferroelectric $K_{1-x}Li_xTaO_3$ (KLT) in the vicinity of its weakly first order transition. This phenomenon is shown to be conceptually similar to the electro-magnetically induced transparency (EIT) phenomenon observed in atomic vapors, which can be modeled classically by a driven master - slave oscillator system. In KLT, it reveals the presence of macroscopic hetero-phase fluctuations and provides unique information on the nature and mechanism of the phase transition in relaxors.

Phase transitions in mixed or partially disordered solids is an area of condensed matter physics in which no single model is applicable to all systems, due to the fact that there is no one single type of disorder. In trying to understand phase transitions in disordered solids, it is therefore particularly useful to find a family of such solids in which the nature of the disorder is well characterized and for which a relatively general phase transition model can be developed. Relaxor ferroelectrics represent such a family, with substitutional disorder. The present Letter reports *a remarkable phenomenon* observed in several relaxor crystals of $K_{1-x}Li_xTaO_3$ (KLT) (but also in $PbZn_{1/2}Nb_{1/2}O_3$-$PbTiO_3$ PZN-PT), which reveals the existence of macroscopic hetero-phase fluctuations [1] and sheds light on the nature and mechanism of the phase transition in these systems.

Relaxor ferroelectrics are highly polarizable systems in which one or several atoms in the unit cell are off-centered and form a dipole that can reorient between several crystallographically equivalent directions. Interactions between off-center ions in the highly polarizable lattice result in the formation of lower symmetry polar nanodomains (PND) [2] the size of which can be estimated through neutron and x-ray elastic diffuse scattering. [3] PNDs can themselves reorient/relax between several equivalent directions, giving rise to the characteristic dielectric dispersion of relaxors. In KLT, lithium ions are off-centered in a cubic direction by almost 1 Å from their normal crystallographic site. [4] For the particular crystal studied (KLT3.5%), the PNDs reach 40 Å in size near the phase transition [3] (~20 Li/PND). Because PNDs are associated with local strain fields, they also form an elastic quadrupole and can mediate a strong coupling between polarization and strain [5] as demonstrated by the observation of dielectric or electro-mechanical (EM) resonances. [6] At lower temperatures, certain relaxors such as KLT, PZN-PT and PSN (Sc) undergo a transition marked by an abrupt drop of their dielectric constant, while the dielectric constant of other relaxors such as PZN or PMN (Mg) decreases smoothly, showing no sign of a transition. [7] The transition in the first group appears to be of the first order and often displays a narrow thermal hysteresis when measured while sweeping temperature continuously. [8,9] However, neither the nature and mechanism of this transition, often crudely characterized as mixed displacive/order-disorder, nor the nature of the low temperature phase are well understood; in KLT for instance, the higher temperature phase is on average cubic, yet characterized by the presence of tetragonal PNDs which represent nuclei of a ferroelectric phase, and the low temperature phase is structurally tetragonal, yet not ferroelectric but either anti-ferroelectric, disordered ferroelectric or a dipolar glass phase with no net spontaneous polarization when cooled in zero field.[10, 11] In the present Letter, we report new measurements of the dielectric resonances in KLT and PZN-PT in the vicinity of the transition, which reveal an unexpected

"transparency window". Analysis of this phenomenon provides new information on the mechanism of the transition in these systems and suggests a possible answer to the above question of the existence and nature of a transition.

The crystal used in the present work was a $K_{1-x}Li_xTaO_3$ (KLT) crystal with 3.5% Li in the form of a bar of dimensions 5 x 5.5 x 8.5 mm with cubic faces. The two main [001] surfaces were fully electroded with evaporated aluminum. Fig.1 shows the dielectric constant measured upon cooling as a function of temperature and at several frequencies. A few points were also taken upon warming and found to fall exactly on the cooling curve (thermal reversibility). In this and the other measurements described in this paper, the temperature was thoroughly equilibrated for approximately 15-20 mn before each measurement, until the dielectric constant was found to be constant in time. The relaxation peak shown in Fig.1 corresponds to the relaxation (90° or π/2 reorientation) of the PNDs between equivalent directions under the effect of an external ac field. [12] As expected for a relaxation peak, it shifts to lower temperature for lower frequencies (relaxor behavior). The transition to the tetragonal phase is marked by a sharp drop in the dielectric constant at $T_c \approx 47$ K, which is more noticeable the lower the frequency. The inset shows birefringence measurements taken under continuous cooling and warming (no equilibration) in a KLT crystal with 3.4% lithium. [13] Instead of a single curve, these reveal a narrow thermal hysteresis loop (see also Ref.9) which signals a two-phase region around the transition. In the vicinity of the transition, local fluctuations between these two phases can therefore be expected. [13, 14]

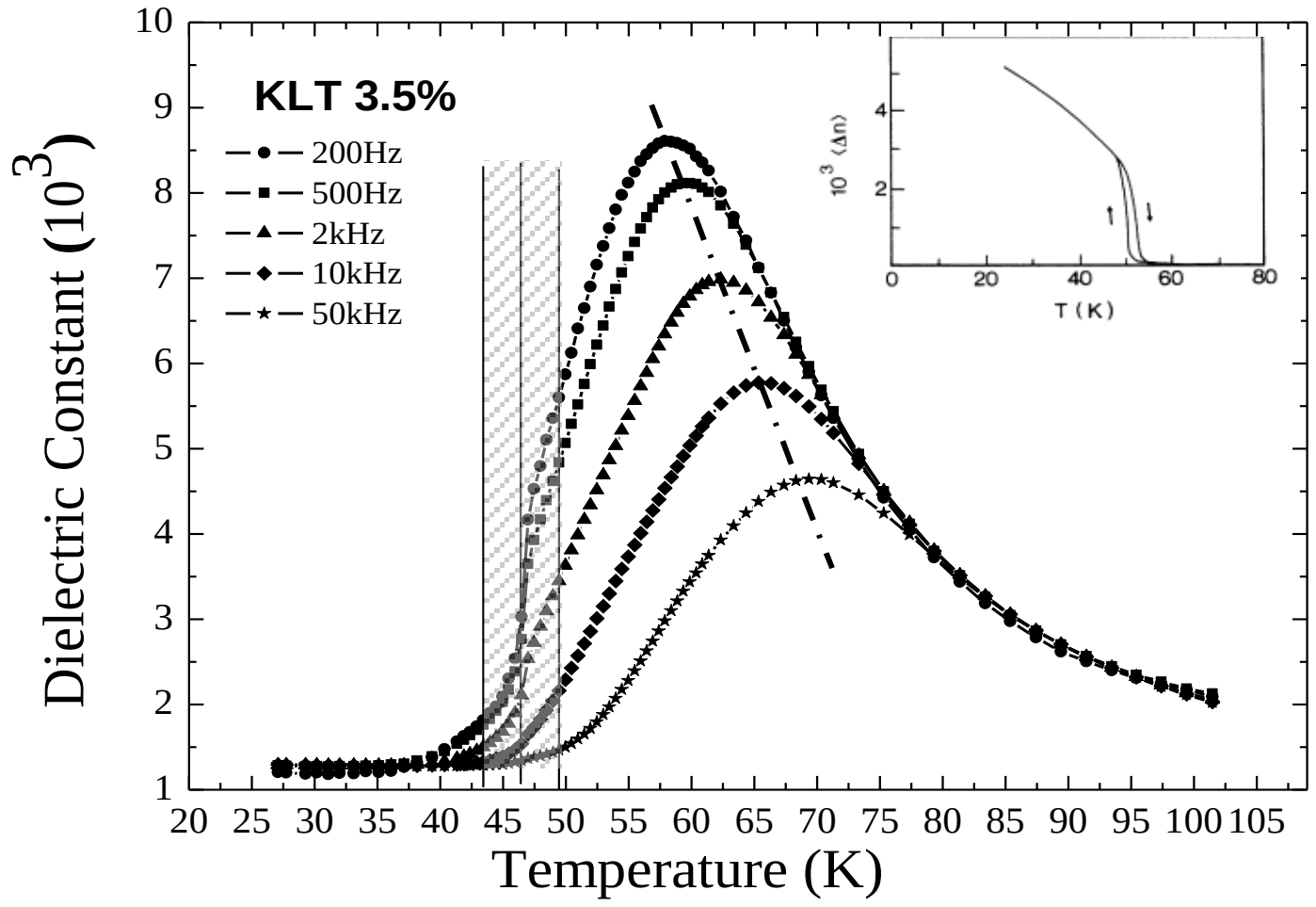

***Fig.1 Real part of the dielectric constant of KLT3.5%measured at different frequencies (inset: birefringence of a different KLT3.4% crystal [Ref. 9])***

In the presence of an external dc bias field, the PNDs are partially aligned, inducing a macroscopic polarization which electrostrictively couples to strain and giving rise to the resonance. At high temperatures (T≥80K), this resonance is narrow (~2kHz) and the relaxation corresponding to the reorientation of the PNDs occurs at much higher frequencies. As the temperature decreases, the relaxation progressively moves to lower frequencies (relaxor behavior in Fig.1). At intermediate temperatures (T=75K-55K), it overlaps with the resonance, which is then overdamped. At lower temperature(T≤55K), once the relaxation has moved to lower frequencies, the resonance reappears, this time exhibiting *a surprisingly sharp and narrow central dip or "transparency window"* in the dielectric resonant absorption peak. Finally, below T≈42K, this sharp and narrow central dip vanishes and the resonance peak recovers it normal shape, as seen later in Fig.6.

Three essential features of the results must be noted:

-i) the dip in the absorption spectrum is only observed in the presence of a dc field

-ii) it is only observed once the relaxation frequency has passed below the resonance frequency, i.e when the reorientation of the PNDs is frozen on the time scale of the resonance

-iii) it is only observed in a temperature range of $\Delta T \approx 7$ degrees centered on the transition, as indicated by the cross-hatched area in Fig.1.

The real and imaginary parts of the dielectric constant of KLT3.5% are shown in greater details at T=44K in Fig. 3a). Because it provides a clue for understanding the KLT results, we also show in Fig.3b the real and imaginary parts of the optical susceptibility of an atomic vapor excited by optical fields [15] and whose energy levels are shown in the inset. The horizontal scale in Fig.3a represents the frequency of the ac field exciting the electromechanical resonance and in Fig.3b, the frequency of the probe laser. The similarity in behavior of the dielectric susceptibility in KLT and the optical susceptibility (solid curve) in the atomic vapor is striking and suggests that similar mechanisms are operating in the two cases. This in no way implies that the two phenomena are identical or physically related, but only that their underlying mechanisms are conceptually similar.

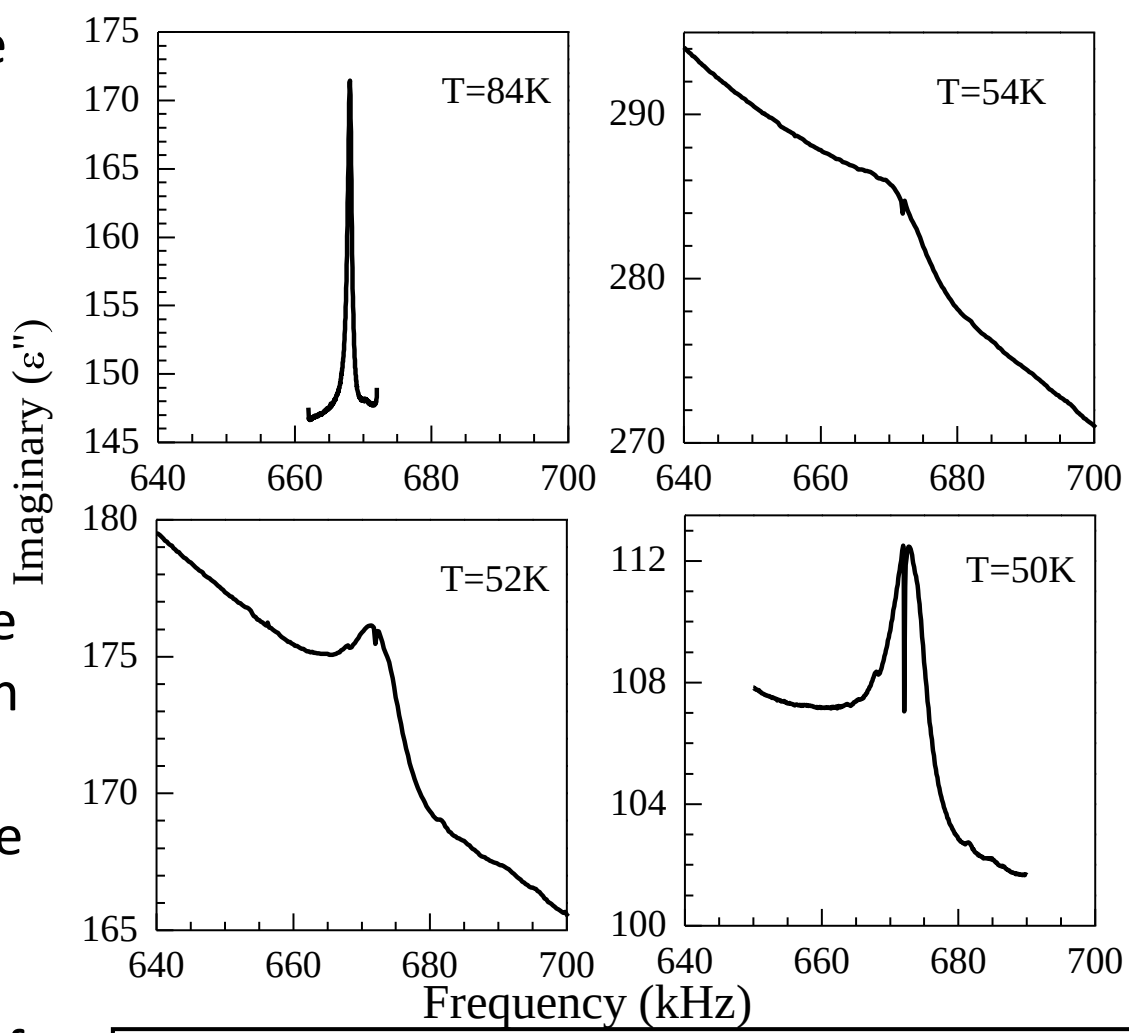

***Fig.2 Imaginary part of the dielectric constant of $K_{1-x}Li_xTaO_3$. The sloped background at the three lower temperatures corresponds to the tail of the broad relaxation peak.***

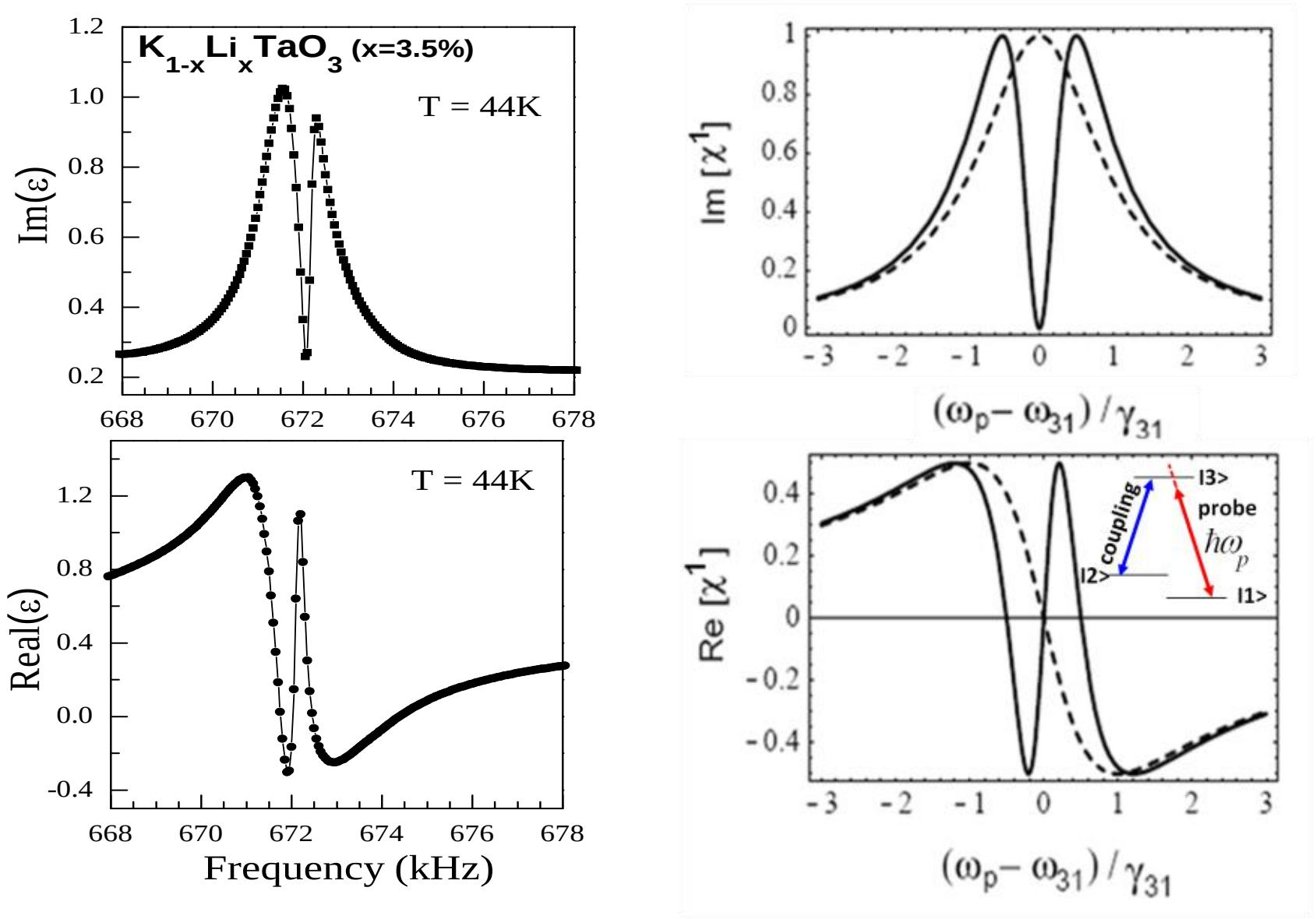

***Fig.3 Imaginary and Real part a) of the dielectric constant of $K_{1-x}Li_xTaO_3$; b) of the susceptibility of an atomic vapor as measured by the probe beam (the dash line is the resonance of a normal radiatively broadened two-level system without EIT)***

*Physical Explanation*

The phenomenon displayed in Fig.3b is the now well-understood electro-magnetically induced transparency or EIT. [16] It gets its name from the fact that, in the presence of a strong coupling beam which puts the atoms in a superposition of states $|2\rangle$ and $|3\rangle$, the on-resonance probe absorption ($\omega_p = \omega_{31}$) goes to zero and the vapor becomes transparent to the probe beam. The origin of this transparency is the destructive interference between *two coherent paths* to the excited state $|3\rangle$: a direct path, $|1\rangle \rightarrow |3\rangle$ and an indirect one $|1\rangle \rightarrow |3\rangle \rightarrow |2\rangle \rightarrow |3\rangle$. [16] In the relaxor ferroelectric case, the probe oscillation is clearly the electro-mechanical one and provides the direct path. The fact that the EIT-like phenomenon is only observed near the phase transition, within a two-phase region, suggests that the coupling oscillations is the "hetero-phase" ones, between the high temperature average cubic phase (with PNDs) and the low temperature homogeneous tetragonal phase.[1, 13] We further develop this point later in the paper.

*Modeling*

Several years ago, Garrido Alzar et al [17] showed that the same behavior was exhibited by a purely classical driven master oscillator-slave oscillator system, which we therefore use to fit our results. Two equivalent versions of such a system are shown in Fig.5: a) with two masses, each attached to their own spring, weakly coupled to each other by a third spring and with only the first mass being driven or b) two RLC circuits, weakly coupled to each other by a coupling capacitor and with only the first circuit being driven. Despite their surprising simplicity, these two realizations contain the essential characteristics that give rise to the EIT-like behavior of the relaxor ferroelectric under study. Although both systems are equivalent, here we use the equivalent circuit model since an RLC circuit is commonly used to describe a piezoelectric (or electrostrictive) resonator (inductor, capacitor and resistor representing respectively the mass or inertia, compliance and damping of the resonator).

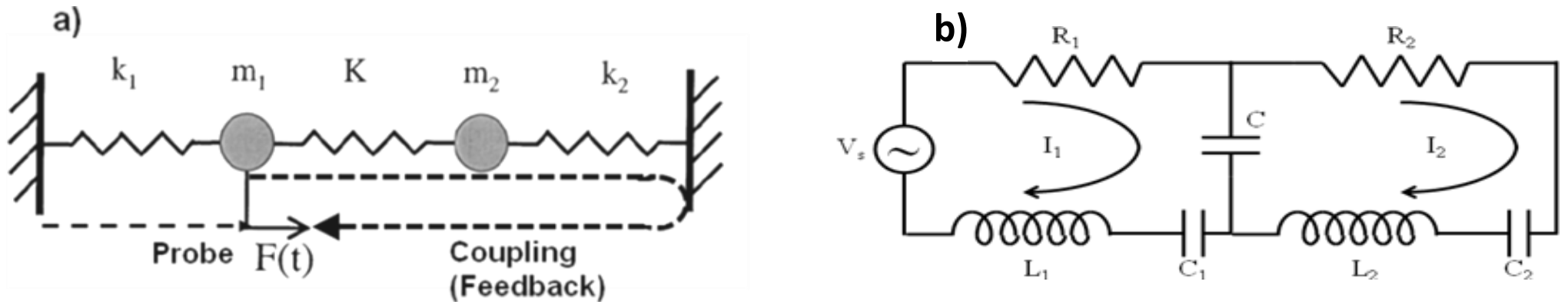

***Fig.5 a) Mass-Spring model and b) Equivalent RLC circuit, both exhibiting EIT-like behavior. Model b) is used to reproduce the dielectric results***

In the relaxor case, the first or driven master oscillator accounts for the electro- mechanical (EM) resonance. The second oscillator is only present in the vicinity of the phase transition where it accounts for the hetero-phase oscillations. The two oscillators are coupled but only the first one (probe) is driven. The capacitor *C* couples the EM and hetero-phase oscillations and ensures coherence between the two paths of the driven excitation: a direct excitation of $(RLC)_1$ and an indirect one via the hetero-phase oscillation represented by $(RLC)_2$. For C infinite, the middle bridge section of the circuit corresponds to a short, the second oscillator is irrelevant and the absorption peak has its normal shape. When the coupling is weak (large *C*), the π phase change at resonance results in destructive interference between the two paths and a sharp dip is observed in the middle of the resonant absorption peak. The expression for the power dissipated in the driven probe oscillator $(RLC)_1$ is easily derived from coupled equations of motion as: [17]

$$P=\frac{p_2}{p_1^2+p_2^2}V_s^2 \text{ where } p_1=\left(X_{1e}-\frac{X_{2e}X_C^2}{|Z_2|^2}\right) \text{ and } p_2=\left(R_1+\frac{R_2X_C^2}{|Z_2|^2}\right) \text{ with } Z_2=[R_2+j(X_{L2}-X_{C2e})]$$

In this expression, $V_s$ is the applied voltage, R is the resistance, *Z* and *X* designate impedances and reactances respectively (the subscripts *L* and *C* indicate inductive or capacitive components and the subscript *e* designates a quantity that includes two capacitances in series, $C_1$ and *C* or $C_2$ and *C*). The experimental and fitted spectra are shown in Fig.6 at three temperatures. A sharp and narrow dip in absorption is clearly seen between 51K and 43K. At 42K, the dip has disappeared and the resonance peak has recovered its normal shape. In practice, given the seven parameters in the fit, the two wingsof the fit, the two wings of the experimental curves were first fitted by a single driven oscillator model, and the parameters obtained were then used as initial parameters for the two-oscillator fit. The quality of the fits is excellent, except in the wings of the peak at 50K, possibly due a residual contribution from the relaxation process seen as a broad background in Fig.2 and subtracted before fitting in Fig.6. The fits at 48K and 46K are not shown for lack of space but are just as good as the fit at 44K shown here. Hence, the goodness of fit further validates the model used to describe the EIT-like phenomenon in KLT.

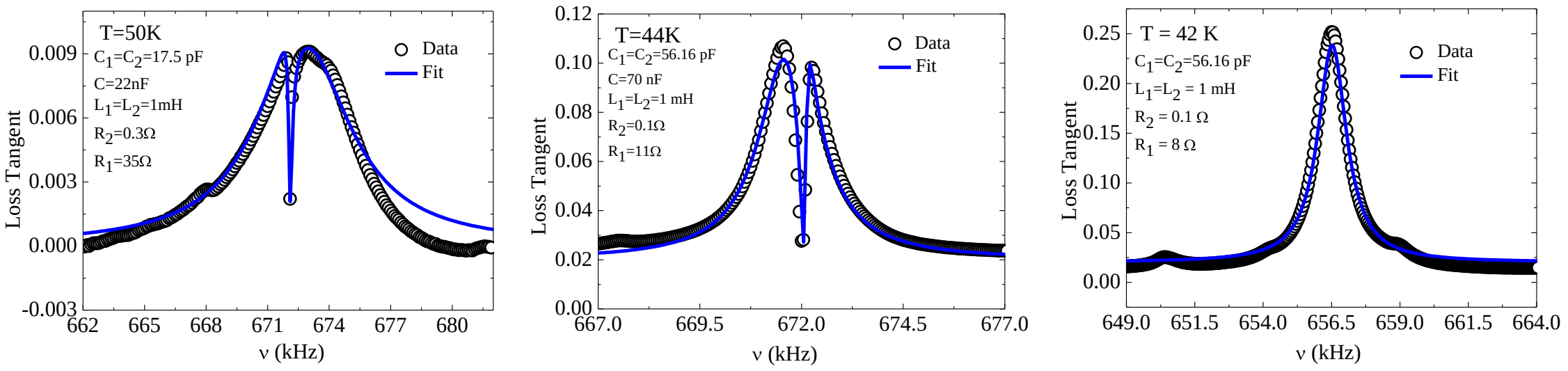

***Fig.6 Imaginary part of the dielectric constant and fit to the two coherently coupled oscillator model***

In Fig. 7, we examine the temperature evolution of the essential fitting parameters, obtained using the RLC model. Since the most physically significant parameters are the capacitances (inverse compliances and coupling), the fitting was done assuming for simplicity $L_1=L_2\approx$ 1μH independent of temperature and $R_2\approx 0.1\Omega$. The capacitances $C_1$ and $C_2$ are found to increase in a stepwise fashion with an inflection point near the transition temperature, $T_c\approx$47K. This reflects a step decrease in the elastic compliance or stiffening of the lattice upon going through the transition and is consistent with a corresponding step increase in the elastic modulus revealed by ultrasonic measurements.[18] It is also consistent with the lower resonance frequency at 42K in the tetragonal phase (Fig.6). The coupling capacitance, $C$, is large (approximately same as $C_1$ and $C_2$) and further increases by a factor of 3 between 50K and 44K. As mentioned earlier, a large C corresponds to the weak coupling case between the EM and hetero-phase oscillations, which explains why the latter are only weakly damped ($R_2 << R_1$) and permits the observation of the coherent interference effect. The resistance $R_1$ decreases by the same factor through the transition, giving a narrower resonance peak at 42K in Fig.6.

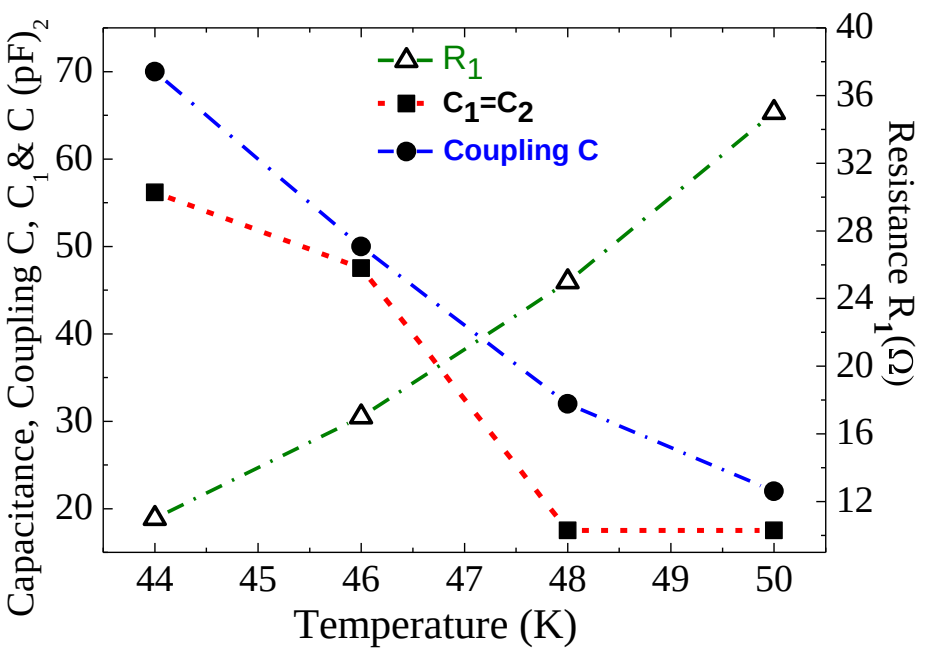

***Fig.7 Temperature evolution of the fitting parameters from the two coupled RLC oscillator model***

The above results call for several remarks:

1) the hysteresis loop shown in the inset of Fig.1 around the transition (measurements made under continuous cooling/warming) covers the very same temperature range in which the EIT-like behavior is observed in our measurements (made in full thermal equilibrium). This clearly indicates that the EIT-like behavior is not simply a characteristic of the PND phase but is related to the coexistence of two thermodynamic phases, here the high temperature average cubic phase and the low temperature tetragonal phase. Critical fluctuations are known to exist near second order but usually not near first order phase transitions. Relaxors present a different case in which PNDs represent stable nuclei of the lower temperature phase, which can trigger critical fluctuations between the high and low temperature phases in the near vicinity of the transition. Within the two phase region, the relative volume fractions of the two phases then fluctuate around their thermal equilibrium value. These fluctuations can also be driven by an appropriate external force. In the present case, if either polarization or strain is the order parameter, the polarization or strain produced by the EM resonance will modulate these relative volume fractions and thereby excite hetero-phase oscillations. As was mentioned earlier, the EIT-like phenomenon is only observed near the transition, when the π/2 relaxation of the PNDs has become frozen on the time scale of the resonance, i.e. when the PNDs can no longer align their polar moments. We must therefore conclude that the hetero-phase oscillations and the transition in KLT are driven by strain and not by polarization. The PNDs therefore act as essential polarization-strain relays inducing the structural transformation. This is consistent with the fact that, while the high temperature phase containing PNDs is cubic on average, the low temperature phase of KLT is structurally homogeneous tetragonal but not ferroelectric. [10] This ordering process is reminiscent of the process proposed by Tagantsev et al. [19] in PMN subjected to a large dc and ac electric field, even though the cause of the ordering is different in the two cases. In both cases, KLT below the relaxor peak and near the phase transition and PMN under a large dc field, the polarization of the PNDs can no longer reorient during the period of the ac field. Instead, the ordering process can be attributed to the sideways motion of the interphase boundaries of the polar nanodomains. In both cases, this motion is a result of the electrostrictive strain resulting from the electric field. This model was in fact used earlier to explain the dynamic response of PNDs in KLT [20], although not in relation with the phase transition.

2) the existence of hetero-phase oscillations near the transition and the step in compliance are both consistent with the presence of tetragonal PNDs as nuclei of the low temperature phase and the first order character of the transition. In a first order transition, the Gibbs free energies of the high and low temperature phases are equal at $T_c$, here $G_{cubic}=G_{tetra}$. Consequently, very little energy is required to excite hetero- phase oscillations near the transition: $dG(X, T_c) = sXdX + pSdX \approx 0$, where $s$ is the elastic compliance, $X$ the stress and $p$ the piezo-entropic coefficient. Taking the second derivative of $G$ with respect to $X$ yields the effective compliance, $s^* = s\left[1 + p\left(\Delta S/\Delta x\Big|_{T_c}\right)\right]$, in which $\Delta x$ is the strain associated with the cubic-tetragonal transition and $\Delta S = S_{cubic}-S_{tetra}$ the entropy discontinuity at a first order transition. The second term in parentheses represents the step observed in the compliance at the transition, which provides a qualitative verification of the experimental results.

Finally, in Fig.8 we also show the EIT-like behavior observed on a PZN-4.5%PT crystal, the dielectric constant of which also exhibits a sharp drop at the transition, as in KLT. This behavior may therefore be common to the subgroup of relaxors that undergo a weakly first order transition driven by strain.

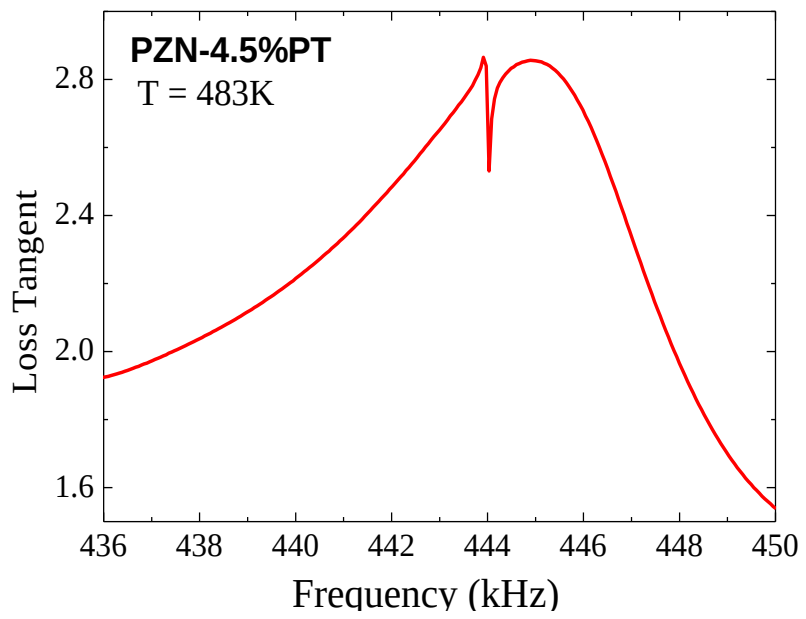

***Fig.8 Loss tangent of PZN-4.5%PT showing an EIT-like dip (crystal from Z.-G. Ye)***

In conclusion, we have reported the observation of “dielectric transparency” in the relaxor ferroelectrics $K_{1-x}Li_xTaO_3$ (KLT) and $PbZn_{1/2}Nb_{1/2}O_3$-$PbTiO_3$ (PZN-PT) excited in EM resonance near their transition. We have shown that this effect is an analog of the electromagnetically-induced transparency (EIT) observed on the optical susceptibility of atomic vapors. In relaxors, this “dielectric transparency” reveals the existence of macroscopic hetero-phase fluctuations within the two-phase region near the transition, which may be a general feature of this family of systems. Although the transition is (weakly) first order, due to the presence of PNDs it is nevertheless accompanied by fluctuations that are otherwise characteristic of second order phase transitions in homogenous systems. The PNDs act as polarization-strain relays inducing the structural transformation.

This work was supported by a grant from the US Department of Energy, Office of Basic Energy Sciences, DE-FG02-06ER46318. The work at Oak Ridge was also supported by USDOE-BES under contract DE-AC05-00OR22725. Special thanks to J. Gunton for useful discussions and for proofreading the manuscript.

*Present address: Physics Department, Lafayette College, Quad Drive, Easton, PA 18042